\documentclass[aps, prb, twocolumn]{revtex4-2}

\usepackage{cmap}

\usepackage[english]{babel}
\usepackage{amsmath}
\usepackage{bm}

\usepackage{graphicx}
\usepackage[caption=false]{subfig}

\usepackage[bookmarks=false]{hyperref}
\hypersetup{
    colorlinks=true,
    linkcolor=blue,
    citecolor=blue,
    urlcolor=blue
}

\begin{document}

\title{Ferroelectric Polarization Reversal vs Pump Spot Shape}

\author{Veniamin A. Abalmasov}
\email{abalmasov@iae.nsc.ru}

\affiliation{Institute of Automation and Electrometry SB RAS, 630090 Novosibirsk, Russia}

\date{\today}

%\begin{document}

%\begin{tocentry}

%\includegraphics[width=1\textwidth]{fig0.png}

%\end{tocentry}

\begin{abstract}

It has recently been shown that the polarization of a ferroelectric can be strongly affected by light pumping or even transiently reversed by mid-IR pumping within a record short subpicosecond timescale. It was then suggested that the reversal of polarization is most likely prevented by the depolarizing field that appears during the process. To minimize the effect of the depolarizing field, I propose to create an elongated domain by choosing a corresponding pump spot shape. I calculate the polarization dynamics for a particular experimental setup with mid-IR pumping and show that the polarization reversal is possible in this case.

\end{abstract}

\maketitle

The polarization created by elementary dipoles in matter is an inherent property of ferromagnetic and ferroelectric materials, which is largely used for data storage \cite{scott1989, plumer2001}, but may be even more in demand in emerging technologies \cite{kakekhani2015, khan2016, sanna2017, wan2021}. The most straightforward way to control polarization by varying the appropriate magnetic and electric fields is limited in speed to hundreds of picoseconds \cite{li2004}. However, the use of ultrashort electromagnetic pulses can reduce this time by tens or even hundreds of times~\cite{kimel2020}. 

To date, ultrafast polarization reversal using femtosecond laser pumping has been successfully achieved in ferromagnets with an unprecedented switching time of about 20~ps \cite{stupakiewicz2017}. At the same time, the situation in ferroelectrics turned out to be more complicated ~\cite{mishina2021}. The original proposal to switch the polarization by resonant excitation of the ferroelectric mode \cite{fahy1994, qi2009, herchig2014} has not yet been fully realized experimentally \cite{katayama2012, grishunin2017, mishina2018, mishina2018Fe}. It was also shown that optical pumping with excitation of the electronic subsystem \cite{lian2019, paillard2019, michel2021, gu2021} leads only to a temporary decrease in polarization \cite{kuo2017, okimoto2017, brekhov2018, burganov2020}. Following the rapid development and achievements of nonlinear phononics \cite{foerst2011, juraschek2017, radaelli2018, juraschek2020}, it was proposed to pump the IR-active optical phonon mode, which is nonlinearly coupled with the ferroelectric soft mode \cite{subedi2015}. However, in this way, only a transient switching of the polarization was achieved in a monodomain LiNbO$_3$ sample, followed by the polarization recovery to its initial value on a timescale of about 0.2~ps~\cite{mankowsky2017}. After that, it was suggested that the polarization reversal was prevented by the depolarizing field created by charges at the domain boundaries and it was proposed to use a metal wire around the pump spot to screen these charges~\cite{abalmasov2020}.

\begin{figure}[t]
\center
\includegraphics[width=0.9 \columnwidth]{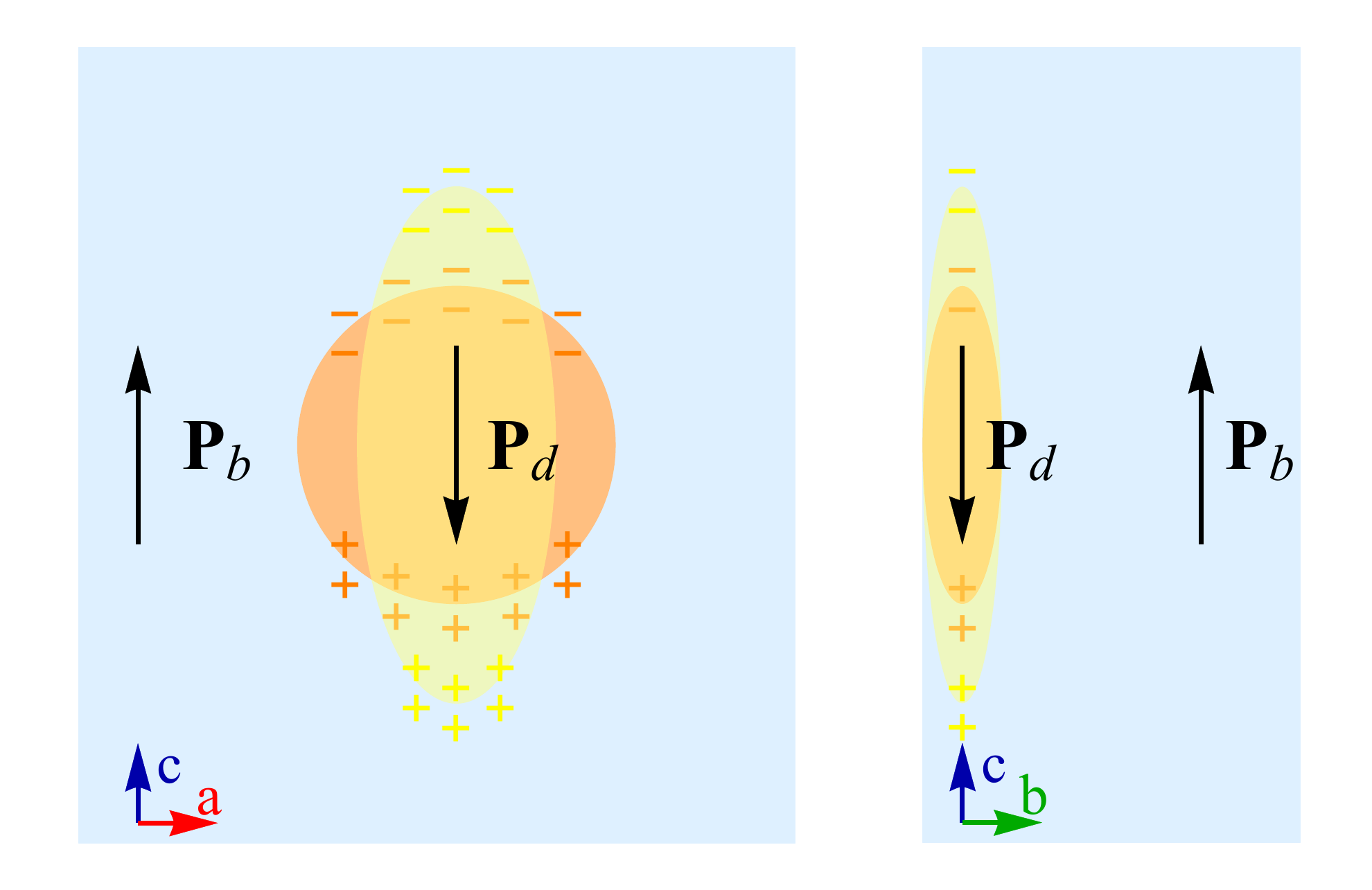}
\caption{Two possible shapes of the pump spot (orange and yellow) and the corresponding domain with opposite polarization and bound charges at its boundaries.}
\label{fig:scheme}
\end{figure}

The influence of the domain shape on the depolarizing field is well known \cite{kittel2004}, and this field, determined by the depolarization factor $N$ for ellipsoidal domains, $E_d = - N P/(\varepsilon_{0} \varepsilon_{\infty})$, where $\varepsilon_{0}$ is the electric constant and $\varepsilon_{\infty}$ is the high-frequency relative permittivity, decreases significantly for elongated domains. Thus, I propose to create an elongated domain by choosing an appropriate pump spot shape in order to reduce the depolarizing field and make polarization reversal possible.

Consider the experimental conditions as in \cite{mankowsky2017}, where in LiNbO$_3$ (LNO) monodomain crystal, the mid-IR resonant pumping of an optical phonon mode nonlinearly coupled to the ferroelectric soft mode led to a transient reversal of polarization. If the domain created in~\cite{mankowsky2017} is approximated by an ellipsoid, the two long axes of which, $a = c$, are equal to the pump spot size of about 65~$\mu$m, the $c$-axis being parallel to the polarization, and the short axis $b$ is equal to the pump penetration depth of about 3.2 $\mu$m (see Fig.~\ref{fig:scheme}), the value of the depolarization factor should be about $N \approx 0.04$ corresponding to the ratio $b/c = 0.05$~\cite{osborn1945}. This value can be further reduced by decreasing the ratio $a/c$. 
For the pump pulse frequency of about 19 THz as in~\cite{mankowsky2017} the smallest possible spot size is about the pump wavelength of 15~$\mu$m. For instance, when the $a$-axis of the spot is halved to $a$ = 30 $\mu$m, and its $c$-axis is doubled to $c = 120$ $\mu$m, keeping the  fluence constant and leading to the ratios $b/c = 0.025$ and $a/c = 0.25$, the depolarization factor is about $N \approx 0.01$, which reduces the depolarizing field by four times~\cite{osborn1945}. 

\begin{figure}[t]
\center
\includegraphics[width=0.98 \columnwidth]{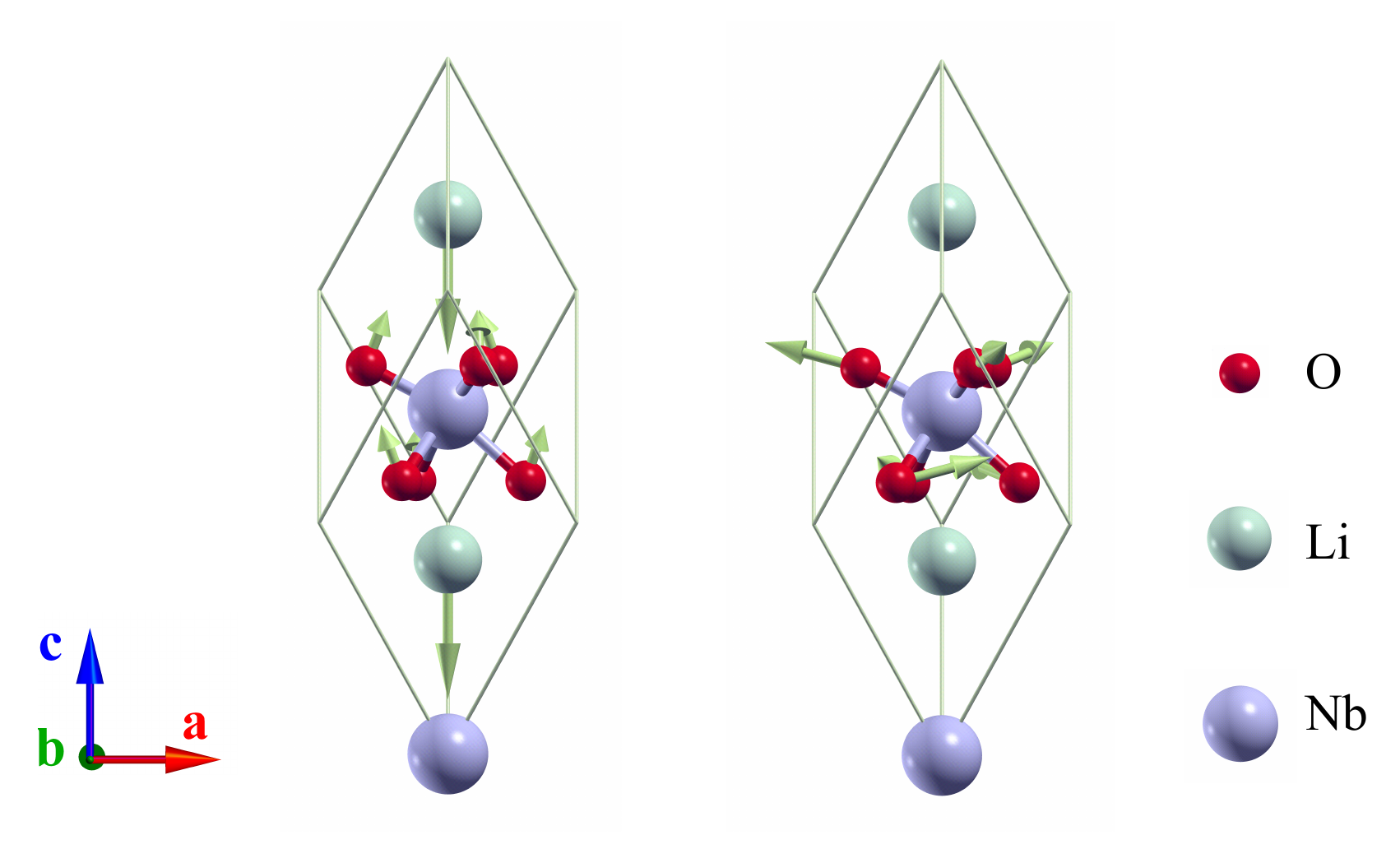}
\caption{LNO primitive unit cell with arrows indicating ion motion in the ferroelectric mode $Q_P$ (left) and the highest frequency infrared-active mode polarized along the c-axis $Q_{\text{IR}}$ (right).}
\label{fig:vectors}
\end{figure}

The LNO crystal in ferroelectric phase (below 1480~K) belongs to $R3c$ symmetry group with 27 optical phonon modes, four $A_1$-modes of which are polarized along the $c$-axis and have a frequency of about 7.5, 8.1, 10, and 19~THz~\cite{kojima2016}.  The calculated {\it ab-initio} using Quantum Espresso~\cite{giannozzi2009, giannozzi2017} IR activities 44.6, 8.8, 0.41, and 53.25~(D/\AA)$^2$/u and frequencies 7.2, 7.7, 9.8, and 18.4~THz of these modes are close to the experimental values~\cite{kojima2016}. The eigenvectors of the first, which is a soft ferroelectric mode, and the last mode, plotted using XCrySDen~\cite{kokalj1999}, are shown in Fig.~\ref{fig:vectors}. The corresponding relative displacements of lithium, niobium and oxygen ions along the $c$-axis for the two modes, hereinafter called $Q_P$ and $Q_{\text{IR}}$, are $-0.563, -0.074, 0.225$ and  $-0.056, -0.049, 0.104$, respectively. The ferroelectric soft mode $Q_P$ involves the motion of lithium and oxygen ions while the high-frequency $Q_{\text{IR}}$ is mainly associated with the motion of oxygen ions in accordance with other {\it ab-initio} calculations~\cite{parlinski2000, postnikov2000, veithen2002, hermet2007, friedrich2015}. However, small displacements of heavy niobium ions are also important because of their large charge of about 6.44 elementary charges for the $zz$-tensor component compared to 1.05 and $-2.53$ for lithium and  oxygen ions, respectively (see also \cite{veithen2002}).

To trace the dependence of the polarization dynamics on the depolarizing field, I solve a system of two equations of motion for the pumped optical phonon coordinate $Q_{\text{IR}}$ and the ferroelectric mode $Q_P$ of the form \cite{abalmasov2020}:
\begin{align}\label{motion}
   \ddot{Q} + \gamma \dot{Q} + \partial F/\partial Q = 0,
\end{align}
where $\gamma$ is the damping constant. 

The Landau free energy in Eq.~(\ref{motion}) can be divided into three parts:
\begin{align}\label{F}
    F = F_0 + F_{E\text{-ph}} + F_{\text{ph-ph}}.
\end{align}
The first part describes free optical phonons:
\begin{align}\label{F0}
    F_0 =  - \frac{\omega^2_P}{4} Q_P^2 + \frac{c_P}{4} Q_P^4 + \frac{\omega_{\text{IR}}^2}{2} Q_{\text{IR}}^2,
\end{align}
where $\omega_{P, {\text{IR}}}$ are the frequencies of the corresponding modes $Q_{P, {\text{IR}}}$. The coefficient $c_P$ determines the equilibrium value $Q_P^e$ through the equation $\partial F/\partial Q_P = 0$. In the absence of electric field and nonlinear phonon interactions this gives $c_P = \omega_P^2/(2 (Q_P^e)^2)$.

The second part of the thermodynamic potential corresponds to the phonon -- electric field coupling:
\begin{align}\label{F_E-ph}
    F_{E\text{-ph}} =   & - E_{p} (Z^*_{Q_P} Q_P + Z^*_{Q_{\text{IR}}} Q_{\text{IR}}) \nonumber \\
    & -  E_d (1 - Q_P/2Q_P^e)) Z^*_{Q_P} Q_P,
\end{align}
where $Z^*_{Q_{P, {\text{IR}}}}$ are the Born effective charges of the corresponding modes $Q_{P, {\text{IR}}}$, $E_d$ is the depolarizing field amplitude, $E_{p}$ is the electric field of the pump pulse directed along the $c$-axis (see Fig.~\ref{fig:scheme}).

\begin{figure}[t]
\center
\includegraphics[width=0.9 \columnwidth]{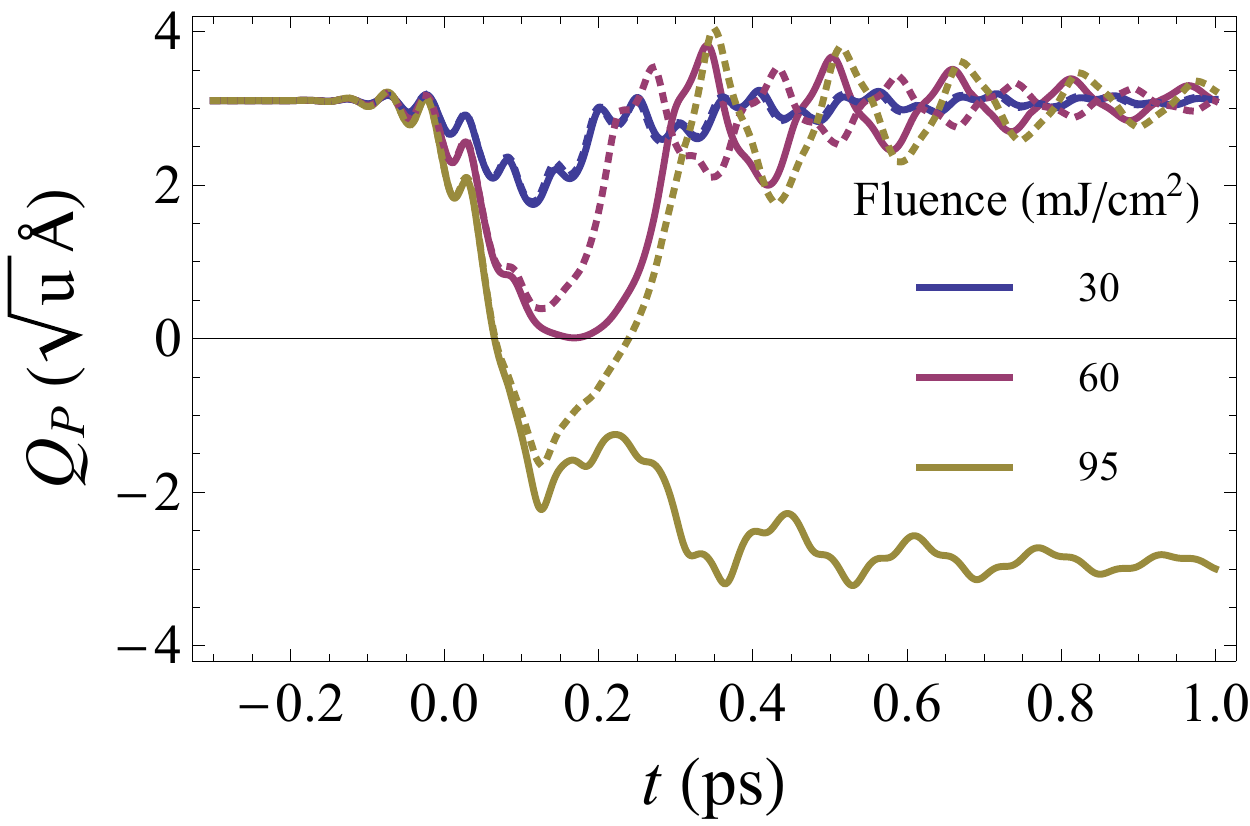}
\caption{Ferroelectric mode dynamics for the pump pulse amplitudes of $E_0$ = 14, 20, and 25~MV/cm when $E_d$ = 2.8~MV/cm (dotted) and $E_d$ = 1.4~MV/cm (solid).}
\label{fig:QP}
\end{figure}

The depolarizing field in a thin slab with polarization normal to the plane is $E_d = P_s/(\varepsilon_{33}\varepsilon_{0})$, which is about 26.4~MV/cm for the saturation polarization $P_s = 0.70$~C/m$^2$ and the dielectric constant $\varepsilon_{33}$ = 30 in LNO~\cite{volk2008}. Assuming that the domain is elliptical, we must multiply this field by the depolarization factor $N$. We should also take into account the fact that for a domain with reverse polarization, the depolarizing field is created by bound charges at the outer boundary of the domain and the inner boundary of the surrounding crystal, and thus doubles. Indeed, a sign change in $Q_P$ in Eq.~(\ref{F_E-ph}) leads to an electric field $2E_d$ when the derivative $\partial F/\partial Q_P$ is considered.

The electric field of the mid-IR pump pulse is $E_p (t) = E_0 \sin(\omega t) \exp(-4\ln2 \,t^2 / T^{2})$ with a carrier frequency $\omega$, a Gaussian envelope with a duration of $T = 0.15$~ps and an amplitude $E_0$ up to 25 MV/cm as in the experiment~\cite{mankowsky2017}.

Finally, the leading phonon coupling terms up to the forth order in phonon amplitudes  are~\cite{abalmasov2020}
\begin{align}\label{F_ph-ph}
    & F_{\text{ph-ph}}  =c_{1} Q_P Q_{\text{IR}}^3 + c_{2} Q_P^{2} Q_{\text{IR}}^{2} +  c_{3} Q_P^{3} Q_{\text{IR}}.
\end{align}

The first term in Eq.~(\ref{F_ph-ph}) does not contribute much in the polarization reversal and was neglected here, while the values of the other coupling constants were taken $c_2 = 34$ meV/u$^{2}${\AA}$^{4}$ and $c_3 = 17$ meV/u$^{2}${\AA}$^{4}$ as in~\cite{abalmasov2020}. Together with the value of the depolarizing field $E_d =2.8$~MV/cm~\cite{abalmasov2020}, this coupling makes it possible to reproduce well the dynamics of polarization observed in~\cite{mankowsky2017} (see dotted lines in Fig.~\ref{fig:QP}). Thus, the depolarization factor turns out to be about $N = 0.1$, which is twice as high as was estimated above from geometric considerations, but nevertheless is in reasonable agreement with all the approximations made. The values of the other parameters used in the calculation were determined in~\cite{abalmasov2020} in accordance with~\cite{mankowsky2017}. The phonon frequencies, dumping constants and Born effective charges of the two modes are $\omega_P = 7.5$~THz, $\omega_{\text{IR}} = 19$~THz, $\gamma_P = 1.6\pi$~THz, $\gamma_{\text{IR}} = 2\pi$~THz and $Z^*_{Q_P} = 1.36$~$e/\sqrt{\text{u}}$, $Z^*_{Q_{\text{IR}}} = 1.40$~$e/\sqrt{\text{u}}$, respectively; the ferroelectric phonon equilibrium value is $Q_P^e = 3.1${\AA}$\sqrt{\text{u}}$.

The polarization dynamics for a half-smaller depolarizing field $E_d = 1.4$~MV/cm, which should be attained for a squeezed pump spot with a ratio $a/c =0.25$, as discussed above, shows polarization reversal (see solid lines in Fig.~\ref{fig:QP}). This reversal becomes possible already for fluences above 60~mJ/cm$^2$.

In experiments with optical pumping \cite{okimoto2017, brekhov2018, burganov2020}, the  penetration depth of the pump pulse varies from about 3~nm \cite{burganov2020} upto 100~nm \cite{brekhov2018}, depending on the crystal used. In this case, the depolarization factor should already be very small, and the depolarizing field is strongly reduced. However, the depolarizing field will decrease even more if the pump spot is elongated, and it would be interesting to see how the spot shape affects the polarization dynamics in this case. In addition, due to the short wavelength of optical pulses, the ratio $a/c$ can be set very small for the same fluence, which would provide an almost zero depolarizing field. I also note that the depolarizing field must be greatly reduced in ferroelectric nanowires \cite{kuo2017, nukala2017}, where, thus, the polarization reversal by the pump pulse should be possible. However, special attention should be paid to the size effects in these samples and their influence on the crystal structure \cite{wang2007, hong2008}.  I note that a $\pi/2$ polarization rotation induced by a strong single-cycle terahertz pulse in some domains of a multi-domain ferroelectric thin film of (Ba$_{0.8}$Sr$_{0.2}$)TiO$_3$, where the depolarized field should be reduced due to the natural shape of the domains, was claimed recently in~\cite{grishunin2017, mishina2018, mishina2018Fe}.

In conclusion, I propose to use an elongated pump spot in experiments with optical, terahertz, or infrared pumping of ferroelectric materials to reduce the depolarizing field in the irradiated domain. I show that this reduction should be sufficient to reverse the ferroelectric polarization using mid-IR pumping in LNO under the conditions of the experiment~\cite{mankowsky2017} on a subpicosecond timescale.

%\bibliography{../sto-pump,../PhaseTransitions, ../BiblioPhononics}


\end{document}